\documentclass{article}
\usepackage{spconf,amsmath,amsfonts,graphicx}
\usepackage{amssymb,textcomp,mathtools}
\usepackage{bm,upgreek,algorithm,hyperref}
\usepackage{multirow,booktabs,hhline,array}
\usepackage{cite,url,makecell,setspace}

\pdfoutput=1

\def\thline{\noalign{\hrule height 1.0pt}}

\renewcommand{\vec}[1]{\bm{\mathrm{#1}}}

\title{End-to-end microphone permutation and number invariant \\ multi-channel speech separation}

\name{Yi Luo$^{\dag}$\sthanks{Work done during internship at Microsoft Research.}, Zhuo Chen$^\ddagger$, Nima Mesgarani$^{\dag}$, Takuya Yoshioka$^\ddagger$}
\address{$^\dag$Department of Electrical Engineering, Columbia University, NY, USA\\$^\ddagger$Microsoft, One Microsoft Way, Redmond, WA, USA}

\begin{document}
\ninept
\maketitle
\setlength{\abovedisplayskip}{2pt}
\setlength{\belowdisplayskip}{2pt}
\setlength{\abovedisplayshortskip}{2pt}
\setlength{\belowdisplayshortskip}{2pt}

\begin{abstract}
An important problem in ad-hoc microphone speech separation is how to guarantee the robustness of a system with respect to the locations and numbers of microphones. The former requires the system to be invariant to different indexing of the microphones with the same locations, while the latter requires the system to be able to process inputs with varying dimensions. Conventional optimization-based beamforming techniques satisfy these requirements by definition, while for deep learning-based end-to-end systems those constraints are not fully addressed. In this paper, we propose transform-average-concatenate (TAC), a simple design paradigm for channel permutation and number invariant multi-channel speech separation. Based on the filter-and-sum network (FaSNet), a recently proposed end-to-end time-domain beamforming system, we show how TAC significantly improves the separation performance across various numbers of microphones in noisy reverberant separation tasks with ad-hoc arrays. Moreover, we show that TAC also significantly improves the separation performance with fixed geometry array configuration, further proving the effectiveness of the proposed paradigm in the general problem of multi-microphone speech separation.
\end{abstract}


\section{Introduction}
\label{sec:intro}
\begin{figure*}[!ht]
	\small
	\centering
	\includegraphics[width=1.6\columnwidth]{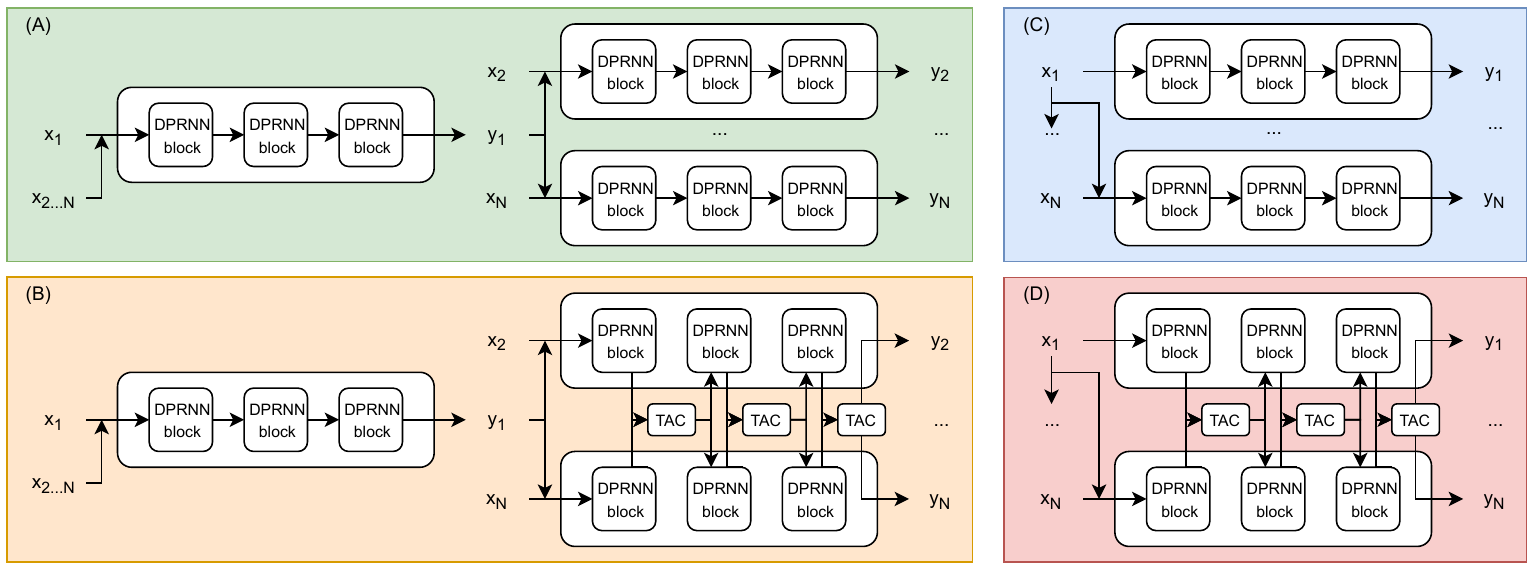}
	\caption{Flowcharts of variants of FaSNet models. Only one output is illustrated for the sake of simplicity. (A) The original two-stage FaSNet. (B) The two-stage FaSNet with TAC applied to every processing block in the second stage. (C) The single-stage FaSNet. (D) The single-stage FaSNet with TAC applied to every processing block.}
	\label{fig:flowchart}
\end{figure*}

Deep learning-based beamforming systems, sometimes called \textit{neural beamformers}, have been an active research topic in recent years \cite{xiao2016study, qian2018deep}. A general pipeline in the design of many recent neural beamformers is to first perform pre-separation on each channel independently, and then apply conventional beamforming techniques such as minimum variance distortionless response beamforming (MVDR) or multi-channel Wiener filtering (MWF) based on the pre-separation outputs \cite{heymann2016neural, erdogan2016improved, xiao2017time, ochiai2017unified, zhang2017speech, heymann2018performance, qian2018deep}. As those conventional beamforming techniques are typically defined as an optimization problem invariant to the permutation of number of the microphones, this pipeline serves as a universal solution to multi-channel speech separation tasks in various configurations. However, as the pre-separation stage is typically trained independently and the estimation of the beamforming filters is a deterministic operation irrelevant to the pre-separation outputs, such systems may generate unreliable outputs when the pre-separation stage fails.

Another pipeline for neural beamformers is to directly estimate the beamforming filters in either time domain or frequency domain \cite{sainath2017multichannel, xiao2016beamforming, meng2017deep, jo2018estimation, luo2019fasnet}. Without the use of conventional filter estimation operations in optimization-based beamformers, this pipeline allows for end-to-end estimation of beamforming filters in a fully-trainable fashion. However, such systems typically assume knowledge about the number of microphones, since a standard network layer can only generate a fix-sized output. Moreover, as the fix-sized output typically consists of the beamforming filters for all channels, it implicitly determines the permutation or indexing of the microphones during the assignment of the sub-parts of the output to different channels. As a consequence, permuting the channel indexes while maintaining their locations might generate completely different outputs and lead to inconsistent performance.

A recently proposed system, the filter-and-sum network (FaSNet) \cite{luo2019fasnet}, attempts to address the disadvantages of both types of pipelines. FaSNet directly estimates the time-domain beamforming filters without specifying the number or permutation of the microphones. With a two-stage design, the first stage applies pre-separation on a selected reference microphone by estimating its beamforming filters, and the second stage estimates the beamforming filters for all remaining microphones based on pair-wise cross-channel features between the pre-separation output and each of the remaining microphones. The filters from both stages are convolved with their corresponding channel waveforms and summed together to form the beamformed output. This is equivalent to replace the filter estimation operation in conventional beamformers by a pair-wise end-to-end filter estimation module and jointly train the two stages. The filter estimation in the second stage is invariant to permutation and number of the microphones due to the use of pair-wise features. Experiment results on fixed geometry array configuration have shown that FaSNet was able to achieve better performance than conventional mask-based neural beamformers in multi-channel speech separation and dereverberation tasks \cite{luo2019fasnet}, indicating the potential of the model.

Although FaSNet overcomes the shortcomings of both pipelines, it also weakens the strengths of them. It still suffers from the problem that the performance of the pre-separation stage greatly affects the filter estimation at the second stage, and the use of pair-wise features prevents it from utilizing the information from all microphones to make a global decision during filter estimation. These flaws might cause unstable and unreliable performance especially in ad-hoc array configurations, where the acoustic properties of different microphones' signals may significantly differ. 

To allow the model to get rid of the weaknesses and preserve the advantages of both pipelines, we propose \textit{transform-average-concatenate (TAC)}, a simple method for microphone permutation and number invariant processing that fully utilizes the information from all microphones. A TAC module first \textit{transforms} each channel's feature with a sub-module shared by all channels, and then the outputs are \textit{averaged} as a global-pooling stage and passed to another sub-module for extra nonlinearity. The corresponding output is then \textit{concatenated} with each of the outputs of the first transformation sub-module and passed to a third sub-module for generating channel-dependent outputs. It is easy to see that, with parameter sharing at the \textit{transform} and \textit{concatenate} stages and the permutation-invariant property of the \textit{average} stage, TAC guarantees channel permutation and number invariant processing and is always able to make global decisions. In Section~\ref{sec:results} we will compare multiple model configurations with and without TAC and show that such design improves the FasNet performance in both ad-hoc and fixed geometry array settings. 

\section{Transform-average-concatenate processing}
\label{sec:model}
\subsection{Tramsform-average-concatenate (TAC)}
\label{sec:TAC-model}
We consider an $N$-channel microphone array with an arbitrary geometry where $N \in \{2,\ldots,N_m\}$ can vary between 2 and a pre-defined maximum number $N_m \geq 2$. Each channel is represented by a sequential feature $\vec{Z}_i \in \mathbb{R}^{T\times *}, i = 1,\ldots,N$ where $T$ denotes the sequence length and $*$ denotes arbitrary feature dimensions. For simplicity we assume one-dimensional features, i.e. $\vec{Z}_i \in \mathbb{R}^{T\times K}$, although the proposed method can be easily extended to higher dimensions. 

A TAC module first transforms each channel's feature with a shared sub-module. Although any neural network architectures can be applied, here we simply use a fully-connected (FC) layer with parametric rectified linear unit (PReLU) activation at each time step:
\begin{align}
    \vec{f}_{i,j} = P(\vec{z}_{i,j}), \, j = 1, \ldots, T
\end{align}
where $\vec{z}_{i, j} \in \mathbb{R}^{1\times K}$ is the $j$-th time step in $\vec{Z}_i$, $P(\cdot)$ is the mapping function defined by the FC layer, and $\vec{f}_{i, j} \in \mathbb{R}^{1\times D}$ denotes the output for channel $i$ at time step $j$. All features $\vec{f}_{i, j}, i=1,\ldots,N$ at time step $j$ are then averaged as a global-pooling stage, and passed to another FC layer with PReLU activation:
\begin{align}
    \hat{\vec{f}}_j = R(\frac{1}{N}\sum_{i=1}^{N} \vec{f}_{i,j})
\end{align}
where $R(\cdot)$ is the mapping function defined by this FC layer and $\hat{\vec{f}}_j \in \mathbb{R}^{1\times D}$ is the output at time step $j$. $\hat{\vec{f}}_j$ is then concatenated with $\vec{f}_{i, j}$ at each channel and passed to a third FC layer with PReLU activation to generate channel-specific output $\vec{g}_{i, j} \in \mathbb{R}^{1\times D}$:
\begin{align}
    \hat{\vec{g}}_{i, j} = S([\vec{f}_{i,j}; \hat{\vec{f}}_j])
\end{align}
where $S(\cdot)$ is the mapping function defined by this FC layer and $[x;y]$ denotes the concatenation operation of vector $x$ and $y$. A residual connection is then added between the original input $\vec{z}_{i,j}$ and $\hat{\vec{g}}_{i, j}$ to form the output of the TAC module:
\begin{align}
    \hat{\vec{z}}_{i, j} = \vec{z}_{i,j} + \hat{\vec{g}}_{i, j}
\end{align}

TAC is closely related to the recent progress in permutation invariant functions and functions defined on sets \cite{zaheer2017deep}. Permutation invariant neural architectures have been widely investigated in problems such as relational reasoning \cite{santoro2017simple}, point-cloud analysis \cite{li2018so} and graph neural networks \cite{xu2018powerful}. The \textit{transform} and \textit{average} stages correspond to the general idea of parameter-sharing and pooling in a family of permutation invariant functions \cite{zaheer2017deep}, while the \textit{concatenate} stage is applied as in the problem setting of beamforming, the dimension of outputs should match that of the inputs. The \textit{concatenate} stage also allows the usage of residual connections, which enables the TAC module to be inserted into any deep architectures without increasing the optimization difficulty.

\subsection{Filter-and-sum network (FaSNet) with TAC}

\subsubsection{FaSNet recap}

Filter-and-sum network (FaSNet) is a time-domain filter-and-sum neural beamformer that directly estimates the beamforming filters with a two-stage design. It first splits the input signals $\vec{x}_i, i=1,\ldots, N$ into frames of $L$ samples with a hop size of $H \in [0, L-1]$ samples:
\begin{align}
	\vec{x}_{i, t} = \vec{x}_i[tH:tH+L-1], \quad t \in \mathbb{Z}
\end{align}
where $t$ is the frame index. Each frame is then concatenated with a context window of $W$ samples in both future and past, resulting in a context frame of $L+2W$ samples:
\begin{align}
	\vec{c}_{i, t} = \vec{x}_i[tH-W:tH+L+W-1]
\end{align}
We drop the frame index $t$ in the following discussions where there is no ambiguity. $C$ beamforming filters of length $2W+1$, $\vec{h}_{i, j}\in \mathbb{R}^{1\times 2W+1}, j=1,\ldots,C$, are estimated from $\vec{c}_i$ for the $C$ target sources, and the waveforms of the sources are obtained by time-domain filter-and-sum operation:
\begin{align}
	\vec{y}_j = \sum_{i=1}^N \vec{h}_{i,j} \circledast \hat{\vec{c}}_i
\end{align}
where $\vec{y}_j \in \mathbb{R}^{1\times L}$ is the beamformed output for source $j$, and $\circledast$ represents the convolution operation. All $\vec{y}_j$ are then converted to waveforms through the overlap-and-add method. 

With a two-stage design, FaSNet first estimates the beamforming filters for a selected reference microphone, which we denote as microphone 1 without the loss of generality. A cross-channel feature of length $2W+1$, which is defined as the normalized cross-correlation feature (NCC), is calculated between $\vec{x}_1$ and each of the $\vec{c}_i, i=1,\ldots,N$:
\begin{align}
	\begin{cases}
		\vec{c}_{i, j} = \vec{c}_i[j:j+L-1] \\
		q_{i,j} = \frac{\vec{x}_1(\vec{c}_{i, j})^T}{\left\|\vec{x}_1\right\|_2 \left\|\vec{c}_{i, j}\right\|_2}
	\end{cases}, \quad j=1, \ldots, 2W+1
\end{align}
It is easy to see that $\vec{q}_i \in \mathbb{R}^{1\times (2W+1)}$ is defined as the cosine similarity between the center frame $\vec{x}_1$ at reference microphone and the context frame $\vec{c}_i$ at microphone $i$ (including the reference microphone itself). A linear layer is also applied on $\vec{c}_1$ to create a $K$-dimensional embedding $\vec{R}_1 \in \mathbb{R}^{1\times K}$ as with the encoder in \cite{luo2019conv}:
\begin{align}
	\vec{R}_1 = \vec{c}_1 \vec{U}
\end{align}
where $\vec{U} \in \mathbb{R}^{(L+2W)\times K}$ is the weight matrix. $\vec{R}_1$ is then concatenated with the mean of all $\vec{q}_i$ and passed to a neural network to calculate the $C$ beamforming filters $\vec{h}_{1,j}, j=1
\ldots,C$. The mean-pooling operation applied on $\vec{q}_i$ guarantees that the cross-channel feature is invariant to the microphone permutations. The beamforming filters $\vec{h}_{1,j}$ are then convolved with $\vec{c}_1$ to generate the pre-separation results $\vec{y}_{1, j} \in \mathbb{R}^{1\times L}$:
\begin{align}
    \vec{y}_{1, j} = \vec{c}_1 \circledast \vec{h}_{1,j},\, j=1,\ldots,C
\end{align}

The second stage of FaSNet estimates source $j$'s beamforming filter for each of the remaining channels based on $\vec{y}_{1, j}$. Similarly, the NCC feature $\vec{v}_{i,j} \in \mathbb{R}^{1\times (2W+1)}$ between $\vec{c}_i, i\in\{2,\ldots,N\}$ and $\vec{y}_{1, j}$ and channel embedding $\vec{R}_i \in \mathbb{R}^{1\times K}$ can be calculated in the same way. Another neural network then takes the concatenation of $\vec{R}_i$ and $\vec{v}_{i,j}$ as input and generates a single beamforming filter $\vec{h}_{i,j}$ for source $j$. All $\vec{h}_{i,j}$ are convolved with their corresponding window $\vec{c}_i$, and summed with the pre-separation output to form the final beamforming output:
\begin{align}
    \vec{y}_j = \vec{y}_{1, j} + \sum_{i=2}^N \vec{c}_i \circledast \vec{h}_{i,j}
\end{align}

\subsubsection{FaSNet variants with TAC}
\label{sec:variants}
The most straightforward way to apply TAC in FaSNet is to replace the pair-wise filter estimation in the second stage to a global operation, allowing the filters for each of the $C$ sources to be jointly estimated across all remaining microphones. For each block in the neural networks for filter estimation, e.g. each temporal convolution network (TCN) in \cite{luo2019conv} or each dual-path RNN (DPRNN) block in \cite{luo2019dual}, the TAC architecture proposed in Section~\ref{sec:TAC-model} is added at the output of each block. Figure~\ref{fig:flowchart} (A) and (B) compare the flowcharts of the original and modified two-stage FaSNet models. 

However, the pre-separation results at the reference microphone still cannot benefit from the TAC operation with the two-stage design. We thus propose a single-stage architecture where the filters for all channels are jointly estimated. Figure~\ref{fig:flowchart} (C) and (D) show the single-stage FaSNet models without and with TAC, respectively. For single-stage models, $\vec{q}_i, i=1,\ldots,N$ is used as the NCC feature for each channel without mean-pooling.

\section{Experimental procedures}
\label{sec:exp}
\begin{table*}[!ht]
	\scriptsize
	\centering
	\caption{Experiment results on ad-hoc array with various numbers of microphones. SI-SNRi is reported on decibel scale.}
	\label{tab:adhoc}
	\begin{tabular}{c|c|c|cccc|c}
		\thline
		\multirow{2}{*}{\thead{Model}} & \multirow{2}{*}{\thead{\# of param.}} & \multirow{2}{*}{\thead{\# of mics}} & \multicolumn{4}{c|}{\thead{Overlap ratio}} & \multirow{2}{*}{\thead{Average}} \\
		\cline{4-7}
		& & & $<$25\% & 25-50\% & 50-75\% & $>$75\% \\
		\hline
		\multicolumn{1}{l|}{TasNet-filter} & 2.9M & \multirow{7}{*}{2 / 4 / 6} & 12.5 / 12.2 / 12.3 & 8.9 / 8.6 / 9.0 & 6.4 / 6.2 / 6.1 & 3.9 / 3.6 / 3.8 & 7.8 / 7.8 / 8.0 \\
		\multicolumn{1}{r|}{+NCC ave.} & 2.9M & & 13.1 / 13.0 / 13.2 & 8.8 / 8.8 / 8.9 & 6.4 / 6.1 / 6.2 & 3.2 / 3.6 / 3.6 & 7.7 / 8.0 / 8.2 \\
		\multicolumn{1}{r|}{\quad+NCC ave.+4ms} & 2.9M & & 13.2 / 13.3 / 13.6 & 9.5 / 9.3 / 9.7 & 7.0 / 6.6 / 7.1 & 4.6 / 4.4 / 4.7 & 8.4 / 8.5 / 9.0 \\
		\cline{1-2}\cline{4-8}
		\multicolumn{1}{l|}{FaSNet} & 3.0M & & 11.0 / 11.5 / 11.5 & 7.0 / 7.9 / 8.1 & 4.5 / 5.2 / 5.4 & 2.0 / 2.6 / 3.0 & 5.9 / 6.9 / 7.3 \\
		\multicolumn{1}{r|}{+TAC} & 3.0M & & 11.3 / 11.8 / 11.7 & 7.2 / 7.8 / 8.5 & 5.1 / 5.4 / 5.5 & 1.9 / 2.3 / 3.0 & 6.2 / 7.0 / 7.4 \\
		\multicolumn{1}{r|}{+joint} & 2.9M & & 14.4 / 13.7 / 14.1 & 10.2 / 9.8 / 10.4 & 7.5 / 7.2 / 7.7 & 4.6 / 4.5 / 4.7 & 9.0 / 8.9 / 9.5 \\
		\multicolumn{1}{r|}{+TAC+joint} & 2.9M & & \textbf{15.2} / \textbf{16.1} / 16.1 & \textbf{10.9} / 11.6 / 12.2 & \textbf{8.6} / 9.5 / \textbf{9.8} & 5.5 / 7.2 / 7.6 & 9.8 / 11.2 / 11.7 \\
		\multicolumn{1}{r|}{\quad+TAC+joint+4ms} & 2.9M & & 15.1 / 16.0 / \textbf{16.2} & 10.8 / \textbf{12.0} / \textbf{12.5} & \textbf{8.6} / \textbf{9.6} / \textbf{9.8} & \textbf{6.2} / \textbf{7.8} / \textbf{8.3} & \textbf{10.0} / \textbf{11.5} / \textbf{12.0} \\
		\thline
	\end{tabular}
\end{table*}

\begin{table*}[!ht]
	\scriptsize
	\centering
	\caption{Experiment results on 6-mic fixed geometry (circular) array. SI-SNRi is reported on decibel scale.}
	\label{tab:fixed}
	\begin{tabular}{c|c|cccc|cccc|c}
		\thline
		\multirow{2}{*}{\thead{Model}} & \multirow{2}{*}{\thead{\# of param.}} & \multicolumn{4}{c|}{\thead{Speaker angle}} & \multicolumn{4}{c|}{\thead{Overlap ratio}} & \multirow{2}{*}{\thead{Average}} \\
		\cline{3-10}
		& & $<$15\textdegree & 15-45\textdegree & 45-90\textdegree & $>$90\textdegree & $<$25\% & 25-50\% & 50-75\% & $>$75\% & \\
		\hline
		\multicolumn{1}{l|}{TasNet-filter} & 2.9M & 7.6 & 7.9 & 8.2 & 8.3 & 12.8 & 9.1 & 6.4 & 3.7 & 8.0 \\
		\multicolumn{1}{r|}{+NCC concat.} & 3.1M & 6.6 & 6.8 & 7.0 & 7.1 & 11.2 & 8.6 & 5.2 & 2.6 & 6.9 \\
		\multicolumn{1}{r|}{+NCC ave.} & 2.9M & 8.2 & 8.6 & 8.9 & 8.9 & 13.3 & 9.9 & 7.1 & 4.4 & 8.7 \\
		\multicolumn{1}{r|}{\quad+NCC ave.+4ms} & 2.9M & 8.5 & 8.8 & 9.1 & 9.3 & 13.6 & 10.0 & 7.3 & 4.8 & 8.9 \\
		\hline
		\multicolumn{1}{l|}{FaSNet} & 3.0M & 8.5 & 9.6 & 10.7 & 11.4 & 14.1 & 11.1 & 8.7 & 6.3 & 10.0 \\
        \multicolumn{1}{r|}{+TAC+joint} & 2.9M & 9.0 & 10.8 & 12.3 & 13.1 & 15.5 & 12.2 & 9.9 & 7.6 & 11.3 \\
        \multicolumn{1}{r|}{\quad+TAC+joint+4ms} & 2.9M & \bf{9.1} & \bf{11.1} & \bf{12.6} & \bf{13.4} & \bf{15.6} & \bf{12.4} & \bf{10.1} & \bf{8.0} & \bf{11.5} \\
		\thline
	\end{tabular}
\end{table*}

\subsection{Dataset}

We evaluate our approach on the task of multi-channel two-speaker noisy speech separation with both ad-hoc and fixed geometry microphone arrays. We create a multi-channel noisy reverberant dataset with 20000, 5000 and 3000 4-second long utterances from the Librispeech dataset \cite{panayotov2015librispeech}. Two speakers and one nonspeech noise are randomly selected from the 100-hour Librispeech dataset and the 100 Nonspeech Corpus \cite{web100nonspeech}, respectively. An overlap ratio between the two speakers is uniformly sampled between 0\% and 100\% such that the average overlap ratio across the dataset is 50\%. The two speech signals are then shifted accordingly and rescaled to a random relative SNR between 0 and 5 dB. The noise is repeated if its length is smaller than 4 seconds, and the relative SNR between the power of the sum of the two clean speech signals and the noise is randomly sampled between 10 and 20 dB. The transformed signals are then convolved with room impulse responses generated by the image method \cite{allen1979image} using the gpuRIR toolbox \cite{diaz2018gpurir}. The length and width of the room are randomly sampled between 3 and 10 meters, and the height is randomly sampled between 2.5 and 4 meters. The reverberation time (T60) is randomly sampled between 0.1 and 0.5 seconds. The echoic signals are summed to create the mixture for each microphone. All microphone, speaker and noise locations in the ad-hoc array dataset are randomly sampled to be at least 0.5~m away from the room walls. In the fixed geometry array dataset, the microphone center is first sampled and then 6 microphones are evenly distributed around a circle with diameter of 10~cm. The speaker locations are then sampled such that the average speaker angle with respect to the microphone center is uniformly distributed between 0 and 180 degrees. The noise location is sampled without further constraints. The ad-hoc array dataset contains utterances with 2 to 6 microphones, where the number of utterances for each microphone configuration is set equal.

\subsection{Model configurations}

We compare multiple models in both configurations. For single-channel models, we use the first stage in the original FaSNet as a modification to the time-domain audio separation network \cite{luo2019conv}, where the separation is done by estimating filters for each context frame in the mixture instead of masking matrices on a generated front-end. We refer to this model as \textit{TasNet-filter}. For adding NCC features to the single-channel baseline, we apply three strategies: (1) no NCC feature (pure single-channel processing), (2) concatenate the mean-pooled NCC features (i.e. first stage in FaSNet), and (3) concatenate all NCC features according to microphone indexes (similar to \cite{gu2019end}, only applicable in fixed geometry array). For multi-channel models, we use the four variants of FaSNet introduced in Section~\ref{sec:variants}. We use DPRNN blocks \cite{luo2019dual} as shown in Figure~\ref{fig:flowchart} in all models, as it has shown that DPRNN was able to outperform the previously proposed temporal convolutional network (TCN) with a significantly smaller model size \cite{luo2019dual}. All models are trained to minimize negative scale-invariant SNR (SI-SNR) \cite{le2019sdr} with utterance-level permutation invariant training (uPIT) \cite{kolbaek2017multitalker}. The training target is always the reverberant clean speech signals. We report SI-SNR improvement (SI-SNRi) as the separation performance metric.

The context window size $W$ is always set to 16 ms (i.e. 256 samples at 16k Hz sample rate) and by default we set $L=W$. We also investigate smaller $L$ while keeping $W$ at 16 ms, which allows us to investigate the effect of window size with dimension of both beamforming filters and the NCC features unchanged (both $2W+1$). The details about the dataset generation as well as model configurations is available online\footnote{\url{https://github.com/yluo42/TAC}}.

\section{Results and discussions}
\label{sec:results}
Table~\ref{tab:adhoc} shows the experiment results on the ad-hoc array configuration. We only report the results on 2, 4 and 6 microphones due to the space limit. For the TasNet-based models, minor performance improvement can be achieved with the averaged NCC features, however increasing the number of microphones does not necessarily improves the performance. For the original two-stage FaSNet models, the performance is worse than TasNet with NCC feature even with TAC applied at the second stage. As TasNet with averaged NCC feature is equivalent to the first stage in the two-stage FaSNet, this observation indicates that the two-stage design cannot perform reliable beamforming at the second stage in the ad-hoc array configuration. On the other hand, single-stage FaSNet without TAC already outperforms both TasNet-based and two-stage FaSNet models, showing that the pre-separation stage is unnecessary in this configuration. Adding TAC to the single-stage FaSNet further improves the performance in all conditions and microphone numbers, and guarantees that more microphones will not make the performance worse. The improvement in conditions where the overlap ratio is high is rather significant. This shows that adding TAC modules enables the model to estimate much better filters by using all available information.

Although TAC is designed for the ad-hoc array configuration where the permutation and the number of microphones are unknown, we also investigate whether improvements can be achieved in a fixed geometry array configuration. Table~\ref{tab:fixed} shows the experiment results with the 6-mic circular array described earlier. We notice that TasNet with all NCC features concatenated leads to even worse performance than the pure single-channel model, indicating that we might need to rethink the properness of feature concatenation in such frameworks. The original FasNet has significantly better performance than all TasNet-based models, which matches the observation in \cite{luo2019fasnet}. However, the single-stage FaSNet with TAC still greatly outperforms the original FaSNet across all conditions, showing that TAC is also helpful for fixed geometry arrays. A possible explanation for this is that TAC is able to learn geometry-dependent information even without explicit geometry-related features.

In the original FaSNet, it was reported that smaller center window size $L$ led to significantly worse performance due to the lack of frequency resolution. Here we argue that the worse performance was actually due to the lack of global processing in filter estimation. In the last row of both tables we can observe better or on par performance for single-stage FaSNet with TAC with 4 ms window. This strengthens our argument and further proves the effectiveness of TAC across various model configurations.

\section{Conclusion}
\label{sec:conclusion}
We proposed \textit{transform-average-concatenate (TAC)}, a simple method for end-to-end microphone permutation and number invariant multi-channel speech separation. A TAC module first \textit{transformed} each input channel feature with a sub-module, and \textit{averaged} the outputs and passed it to another sub-module, finally \textit{concatenated} the output from second stage with each of the output from the first stage and passed it to a third sub-module. The first and third sub-modules were shared for all channels. TAC can be viewed as a function defined on sets being invariant to the permutation and number of set elements, and guaranteed to use the fully information within the set to make global decisions. We showed how TAC can be inserted seamlessly into the filter-and-sum network (FaSNet), a recently proposed end-to-end multi-channel speech separation model, to greatly improve the separation performance in both ad-hoc and fixed geometry configurations. We hope TAC can shed light on model designs for other multi-channel processing problems.

\bibliographystyle{IEEEbib}
\bibliography{refs}

\end{document}